\documentclass[twocolumn,english,aps,prx,floatfix,amssymb,superscriptaddress,longbibliography]{revtex4}
\usepackage[latin9]{inputenc}
\usepackage{verbatim}
\usepackage{float}
\usepackage{amsmath}
\usepackage{amssymb}
\usepackage{graphicx}
\usepackage{color}
\usepackage{xcolor}
\usepackage{tikz}
\newcommand{\ket}[1]{\ensuremath{\left| #1 \right>}}

\usepackage{bbold}
\DeclareMathOperator{\sgn}{sgn}

\usepackage[bookmarks=false,linkcolor=blue,urlcolor=blue,colorlinks,citecolor=blue]{hyperref}
\makeatletter

\newcommand{\be}{\begin{equation}}
\newcommand{\ee}{\end{equation}}
\newcommand{\bea}{\begin{eqnarray}}
\newcommand{\eea}{\end{eqnarray}}

\begin{document}

\title{Rise and fall of Yu-Shiba-Rusinov bound-states  in charge conserving $s$-wave one-dimensional superconductors}

\author{Parameshwar R. Pasnoori}
\affiliation{Department of Physics and Astronomy, Rutgers University, Piscataway, NJ 08854-8019 USA}
\author{Natan Andrei}
\affiliation{Department of Physics and Astronomy, Rutgers University, Piscataway, NJ 08854-8019 USA}
\author{Colin Rylands}
\affiliation{SISSA and INFN, via Bonomea 265,  34136 Trieste ITALY}
\author{Patrick Azaria}
\affiliation{Laboratoire de Physique Th\'eorique de la Mati\`ere Condens\'ee, Sorbonne Universit\'e and CNRS, 4 Place Jussieu, 75252 Paris, FRANCE}

\begin{abstract}
We re-examine  the problem of a magnetic impurity coupled to a superconductor focusing on the role of quantum fluctuations.  We study in detail, a system that consists of  a one-dimensional charge conserving spin-singlet superconductor coupled to a boundary magnetic impurity. 
Our main finding is that quantum fluctuations lead to the destruction of Yu-Shiba-Rusinov (YSR) intra-gap bound-states in all but a narrow region of the phase diagram. We carry out our analysis in three stages, increasing the role of the quantum fluctuations at each stage. First we consider the  limit of a classical impurity and study the bulk semiclassically, finding YSR states throughout the phase diagram, a situation similar to conventional BCS superconductors. In the second stage, we reintroduce quantum fluctuations in the bulk and find that the YSR state is suppressed over half of the phase diagram, existing only around the quantum critical point separating the unscreened and the partially screened phases. In the final stage we solve exactly the full interacting model with arbitrary coupling constants using Bethe Ansatz. We find that including both the quantum fluctuating bulk and quantum impurity destabilizes the YSR
state over most of the phase diagram allowing it to exist only in a small region, the YSR regime, between a Kondo-screened and an unscreened regime. Within the YSR regime a first order phase transition occurs between a spin singlet and doublet ground state. We also find that for large enough impurity spin exchange interaction a renormalized Kondo-screened regime is established. In this  regime, not found for BCS superconductors, there is no YSR state and a renormalized Kondo temperature scale is generated. 
 \end{abstract}
\maketitle
\section{Introduction}
The study of magnetic impurities in superconductors  has a long history~\cite{ Shiba,Yu, Rusinov, Zitt, matsuura,sakurai, balatsky, heinrich, Aoi,Maple,LUENGO}, but still remains of great significance in contemporary condensed matter physics~\cite{ Yazdani,jistm,Zazaunov,demler1,demler2,demler3}. This is due to the extensive recent experimental advancements in controlling a single magnetic adatom on superconducting substrates \cite{Franke,tristanexp,peterexp} and the prospect of realizing  nano-structures with topological character \cite{Yazdani2,Glazman,naoto,pascal}. An important element in this direction   is the existence  of bound-states with energies lying within the superconducting gap,  localized close to the magnetic impurity: the so called Yu-Shiba-Rusinov (YSR) states~\cite{Shiba,Yu, Rusinov}.
These states play a key role in the Kondo screening of the magnetic impurity in BCS type superconductors  and  form the basis of  current understanding for the coupling between
 adatoms  which eventually may lead to one-dimensional topological structures \cite{morari,jiangliang,choy,falkopientka,dloss,bernevigyazdani,heimes,heimes2,brydon,perge,sau1, ruby,dloss2,feldman}.

 \begin{center}
\begin{figure}[!h]
\label{qualitative}
\includegraphics[trim=0 50 0 0, width=0.95\columnwidth]{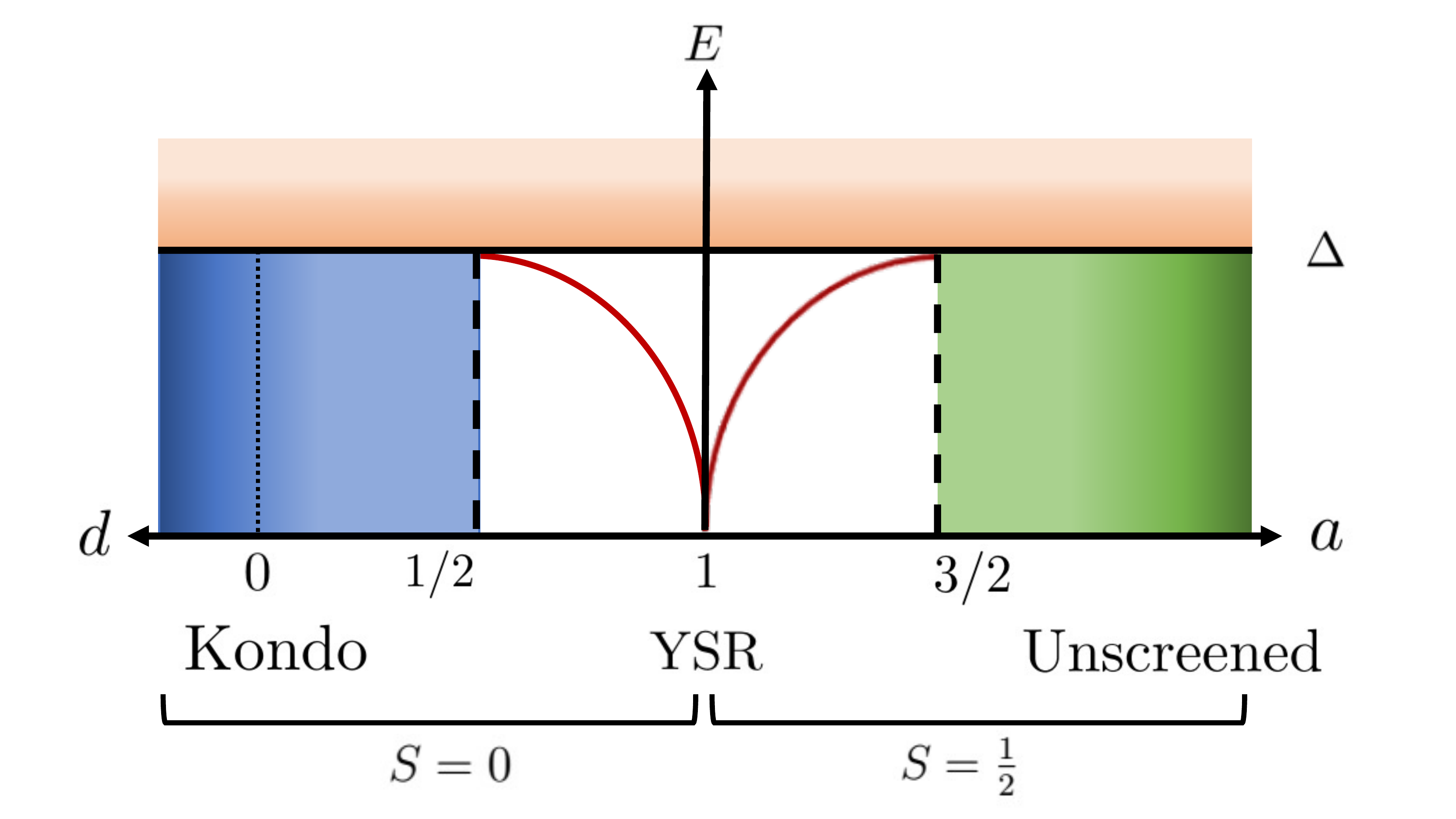}
 \caption{ 
Schematic phase diagram of the Gross-Neveu model with a quantum spin $S=1/2$ impurity at one of its edge. The figure displays   the spectrum as a function of  $d$,     the RG invariant which encodes the bulk and impurity coupling constants,  see equation \eqref{RGinv} (when  $d$ turns imaginary we use $d=ia$). The  YSR state exists only for  $\frac{1}{2}<a<\frac{3}{2}$, its energy is the solid red line. A first order quantum phase transition between screened ($S=0$) and unscreened ($S=\frac{1}{2}$) phases occurs at $a=1$.  For $a>\frac{3}{2}$, the impurity is unscreened and no YSR states exist. In the Kondo regime, $a<\frac{1}{2}$ or $ d \in \mathbb{R}$, a many body screening of the impurity takes place 
 and a renormalized Kondo scale $T_K>\Delta$ is generated.    The shaded orange represents the continuum of bulk states. For comparison, the BCS phase diagram would exhibit
  YSR states  throughout. 
}
\end{figure}
\end{center}
 
In a BCS $s$-wave superconductor the YSR state is a  bound-state 
resulting from the scattering of Bogoliubov quasiparticles  on a {\it classical} spin impurity $S$.
It is  associated with the longitudinal exchange interaction $J S^z s^z$
between the impurity and the bulk electrons where $J$ is the contact exchange coupling,
 $s^z$ is the spin density of the electrons at the impurity site and $S^z=S$.
The YSR state  corresponds to a localized  excited-state electron which  aligns its spin opposite
to that of the impurity and partially screens it. Its energy $\epsilon$ lies below the superconducting 
gap $\Delta$, 
\be
\label{YSRenergy}
\epsilon=\Delta \cos{2\alpha},
\ee
with the phase shift $\alpha = \tan^{-1}( \pi N_0 JS/2)$,
$N_0$ being the normal state DOS at the Fermi level. When $\epsilon > 0$
the impurity is unscreened whereas it is partially screened when $\epsilon < 0$; a quantum
phase transition between these two states taking place at $\epsilon = 0$ when they cross~\cite{sakurai}.

When  a quantum $S=1/2$  impurity is  considered (with the electrons described by a BCS mean field theory),  quantum fluctuations
described by  spin-flip processes  $J (S^+ s^- + S^- s^+)$ need to be  taken into account. These  lead to the Kondo effect in the normal state whereas in the superconductor the accepted view is that they merely renormalize  the exchange interaction between the bulk electrons and the impurity. A localized intra-gap bound-state, 
whose energy is  given by (\ref{YSRenergy}), still exists but with a phase shift $\alpha$
depending on the ratio $T^0_K /\Delta$ where  $T^0_K \propto e^{-\pi/2J}$  is the bare  Kondo temperature
in the normal state.  At the quantum level, the YSR state is seen as an intra-gap state which   arises from the interplay between  
Cooper pairing  and Kondo screening  of the impurity spin by bulk fermions. 
When $\epsilon < 0$ (i.e: $T^0_K /\Delta \gg 1$), Kondo screening takes place  before the onset of superconductivity and the ground-state is a spin singlet with total spin (including both the impurity and electron spins) $S_{\rm T}=0$ and an odd fermion parity~\cite{Zitt}.
When $\epsilon >0$ (i.e: when $T^0_K /\Delta \ll 1$)  the Kondo effect is suppressed by superconductivity: the impurity spin remains unscreened,   the ground-state is a  doublet with spins $S_{\rm T}^z=\pm1/2$ with an even fermion parity~\cite{matsuura}. 
 Results from numerical renormalization group (NRG) indicate that a phase transiton occurs when $T^0_K /\Delta \simeq 0.3$~\cite{SakaiShimizu}.

 
The question we shall address in this work  is how this scheme survives  strong phase fluctuations of the superconducting order parameter.
 This is typically the case of one-dimensional,  $s$-wave charge conserving
superconductors where superconductivity is not induced by  proximity  effect but is intrinsically 
due to the attractive interactions between the electrons. In this case
quantum fluctuations  destroy superconducting  long-range order
leaving the system  with only  quasi long-range superconducting order, i.e.
a spin gap opens (and hence also does  the single particle gap), but Cooper pair correlations functions display only  power-law asymptotics. 
One may hence anticipate that
fluctuations may  strongly affect  the mean field description  of the YSR states. We shall indeed  find in this work  that the strong quantum fluctuations in the bulk significantly  modify the BCS picture described above, both when  a classical impurity or a quantum impurity are considered.
Before going into the details of our study let us first  briefly present our results.

For  a classical impurity we find, by solving  different limits of an anisotropic bulk superconductor model consistent with the classical impurity limit,   an intra-gap YSR bound-state as in the BCS case.   However
it  exists only  in a domain of exchange interaction that  {\it narrows}  as one increases the quantum  fluctuations in the bulk. 
Still, as in the BCS case, the YSR level  controls the phase transition between a partially screened and unscreened phase at a quantum critical point where it has zero energy.
When a quantum $S=1/2$ spin impurity is considered,  the interplay between the spin-flip quantum fluctuations
of the impurity spin with those of the superconducting phase leads to even more striking effects. The qualitative phase diagram of the Gross-Neveu model, which is the  field theory describing a $SU(2)$ invariant one-dimensional s-wave superconductor,  with a spin $S=1/2$
impurity at one of its edges is  depicted   in Figure~\ref{qualitative}.
To begin with, in our exact solution, the control parameter is a Renormalization Group (RG) invariant $d$, which can be real or purely imaginary with $d=ia$, and is a function of the couplings (see Eqs.(\ref{RGinv}, \ref{couplings})) and {\it not} of $T_K^0/\Delta$. 
For moderate exchange interaction, which correspond to the domain $1/2< a< 3/2$, the  spin-flip processes renormalize the exchange interaction as in the BCS case and the YSR state still exists in a narrow part of the phase diagram, thereafter coined YSR regime, between  the  doublet and screened phases found in \cite{PRA}. 
However, when the exchange interaction is large enough, i.e. when $d\in \mathbb{R}$ and $a < 1/2$, 
the YSR level  disappears from the spectrum and a renormalized Kondo screened  regime
is stabilized. In this regime the spin-flip processes are  enhanced by the quantum fluctuations
in the bulk and the screening of the  impurity resembles a genuine Kondo effect.
The Kondo screening in the superconductor is then characterized by a  
{\it renormalized } Kondo temperature $T_K \ge \Delta$ which is  {\it different} from
the bare Kondo temperature $T_K^0$ of  the normal state. For instance deep in the Kondo regime i.e. when  $T_K \gg \Delta$,
the  impurity DOS takes the characteristic  Kondo Lorentzian form $\rho_{\rm imp}(E) \simeq \frac{1}{\pi}  T_K/(E^2 +(T_K)^2)$
and $T_K$ marks the energy scale where the renormalized exchange interaction reach the strong coupling regime
while the bulk pairing interaction remains at weak-couplings.  As $T_K$ decreases towards 
$\Delta$, a smooth cross-over between many-body screening to single particle screening of the impurity takes place. Similarly, for weak enough exchange coupling, i.e. when $a > 3/2$,  the YSR states also disappear from the spectrum. The system is in an unscreened regime, the ground state is a two fold degenerate spin doublet. These results highlight the significant impact that quantum fluctuations can have on the existence of YSR states. Although we have focused on a concrete example of a one-dimensional charge conserving superconductor our findings suggest that the role YSR states  need to be reexamined in other non-conventional superconductors where fluctuations are prominent, including higher dimensionsal superconductors.

In the following we shall present a detailed study of the effect of a magnetic
impurity in one-dimensional charge conserving s-wave superconductors. We shall use a field theoretic approach in which these systems are described by the one-dimensional Gross-Neveu (GN) model and study both the case of a classical impurity and of a quantum spin $S=1/2$ one. 
This paper is organized as follows. In the first section (\ref{Model}), we  introduce our model and recall some properties of one-dimensional charge conserving superconductors. In section (\ref{classicalspin}) we study the case of a classical spin impurity in a suitable anisotropic limit where bulk fluctuations are minimal and also
at a special point where it is described in terms of free fermions (Luther-Emery point).
In section (\ref{bethe}) we consider the case of quantum spin $S=1/2$ impurity. We discuss the Bethe ansatz solution and relegate the details of the calculation to the appendix \ref{Bethe}.  Finally, in section (\ref{discussions}) we discuss our results and comment on possible implications for other systems.

\section{The Gross-Neveu model with a spin $S=1/2$ impurity\label{Model}}

Our approach is a field theoretic one in which we shall model the superconducting bulk by the Gross-Neveu model. The latter model is the well-known effective low- energy field theory describing one dimensional s-wave superconductors such as the attractive Hubbard model in one dimension. In this framework,   the effects of  a magnetic impurity  in a one dimensional 
s-wave superconductor is described by the Hamiltonian 
\be
\label{Hamiltonian}
H= H_{\text{GN} }+ H_{\text{imp}},
\ee
where $H_{\text{GN}}  = \int_{-L/2}^{0} dx\;  \mathcal{H}_{\text{GN}} $
is the Hamiltonian of the Gross-Neveu (GN) model with
\bea
\label{GN}
 \mathcal{H}_{\text{GN}}&=&-i (\psi^{\dagger}_{R a}\partial_x \psi^{}_{R a}-\psi^{\dagger}_{L a}\partial_x \psi^{}_{L a})-  \\
 &-&2g \psi^{\dagger}_{Ra}\psi^{\dagger}_{Lc}\left(\sigma^x_{ab}\sigma^x_{cd}+\sigma^y_{ab}\sigma^y_{cd}+\sigma^z_{ab}\sigma^z_{cd}\right)\psi_{Rb}\psi_{Ld} \nonumber,
\eea
and 
\be
\label{kondo}
H_{\text{imp}}= -J\vec{\sigma}_{ab}\cdot\vec{S}_{\alpha\beta}\psi^{\dagger}_{La}(0)\psi^{}_{Rb}(0),
\ee
where we have set the Fermi velocity $v_F\equiv 1$ and  have  absorbed a factor 
$\pi N_0$   ($N_0$ being the DOS at the Fermi level in the normal state)
 into the  definition of the contact antiferromagnetic exchange interaction $J >0$.
In the above equations $\vec{\sigma}_{ab}, \vec{S}_{\alpha\beta}$ are the Pauli matrices acting in the spin spaces of the bulk fermions  and impurity spin respectively and  $\psi_{L(R) a}(x)$, $a=(\uparrow, \downarrow)$, 
describe left and right moving  fermions carrying spin 1/2. In the geometry we are considering they satisfy open boundary conditions (OBC) at both left and right edges which read, 
\be
 \psi_{R}(-L/2)= - \psi_{L}( -L/2), \;  \psi_{R}(0)= - \psi_{L}( 0).
\label{OBC}
\ee
where we have introduced the spinor notation $\psi_{L(R)}^T \equiv (\psi_{L(R) \uparrow}, \psi_{L(R) \downarrow})$.
The Hamiltonian (\ref{Hamiltonian}) conserves the total number of fermions $N$, 
\be
\label{N}
N=  \int_{-L/2}^0 dx\;  (\psi^{\dagger}_L(x) \psi^{}_L(x) + \psi^{\dagger}_R(x) \psi^{}_R(x)),
\ee
and is spin rotation invariant as it commutes with the total spin operator 
\be
\vec S_{\rm T}= \vec s + \vec S_{},
\ee
where $\vec S $ is the impurity spin operator and $\vec s_{} =  \int_{-L/2}^0 dx\; \vec s_{}(x)$ with
\be
\label{S}
\vec s_{}(x)= \frac{1}{2} (\psi^{\dagger}_L(x) \vec \sigma \psi^{}_L(x) + \psi^{\dagger}_R(x) \vec \sigma \psi^{}_R(x)),
\ee
that of the bulk fermions. In the absence of the four-fermion interaction, i.e: when $g=0$, the Hamiltonian (\ref{Hamiltonian}) describes a Kondo impurity in a metal and its physics is well understood. When $J=0$ and $g > 0$, the GN model describes an s-wave superconductor:
a spin gap $\Delta=D e^{-\pi/2g}$ (where $D$ is a cut-off scale) opens which stabilizes quasi-long range s-wave superconducting order with $\langle {\cal O}_s(x) {\cal O}_s(0) \rangle \sim |x|^{-1/2}$ where ${\cal O}_s \propto \psi^{\dagger}_{R\uparrow}\psi^{\dagger}_{L\downarrow} - \psi^{\dagger}_{R\downarrow}\psi^{\dagger}_{L\uparrow}$. When both $g >0$ and $J>0$ the superconducting instability compete with the Kondo effect and the model (\ref{Hamiltonian}) is well suited to study the impact of
superconductivity on the Kondo effect and vice-versa.

\section{Classical spin impurity\label{classicalspin}}

Let us first  consider the case where the
impurity is  treated as a classical spin. This can be achieved by taking the limit of large quantum spin $S$ together with that of small exchange interaction $J >0$ while keeping $JS={\rm cst}$. In this limit the  localized
impurity spin acts as a local magnetic field, in say the $``z"$ direction, and the Kondo interaction becomes strongly anisotropic
\be
\label{kondoclassical}
  H_{\rm imp}= - JS\;  \left(\psi^{\dagger}_{L} \sigma^z_{}\psi^{}_{R}\right)(0).
  \ee
As we shall now see, in one dimension, the boundary magnetic field  can be reabsorbed into the boundary conditions. Consider indeed the shift  
\bea
\label{transf}
{\psi}_{R(L)}(x)&=&e^{\pm i \alpha(x) \sigma^z} \;{\psi'}_{R(L)}(x),
\eea
with
\bea
 \alpha(x) &=& - \alpha_0 [1+ \sgn(x)], \; \alpha_0 = \frac{JS}{2},
\eea
where  $\sgn(x)$ is  the sign function. Since $\alpha(x)$ vanishes in the bulk (where $x<0$) and due to the OBC (\ref{OBC}),
the  Hamiltonian (\ref{Hamiltonian}) reduces to that of the GN model (\ref{GN})  without any coupling to the impurity spin, i.e:
\be
\label{Hprime}
H(\psi, g, J) = H(\psi',g,0),
\ee
but with {\it twisted} OBC at the right boundary $x=0$ for the shifted fermions 
\bea
\label{twistedOBC}
\psi^{'}_{R}(-L/2) &=& - \psi^{'}_{L}(-L/2), \nonumber \\
\psi^{'}_{R}(0) &=& -e^{i2\alpha_0 \sigma^z} \psi^{'}_{L}(0). 
\eea
Even with this simplification, the problem remains non trivial \cite{PAN2}. To gain some insights we begin by considering an anisotropic bulk 
interaction  for which approximate methods are available.
 
\subsection{Breaking $SU(2)$ symmetry in the bulk}

We shall now   break the $SU(2)$ symmetry in the bulk,  consistently  with the strong anisotropic  Kondo coupling (\ref{kondoclassical}),  and consider  the $XXZ$ deformation of the GN model in which the quartic term in (\ref{GN}) is replaced by
 \bea
\label{U1Thirring}
 \psi^{' \dagger}_{R,a}\psi^{'}_{R,b}\left( g_{\perp} [\sigma^x_{ab}\sigma^x_{cd}+\sigma^y_{ab}\sigma^y_{cd}]+ g_{\parallel} \sigma^z_{ab}\sigma^z_{cd}\right)\psi^{' \dagger}_{L,c}\psi^{'}_{L,d},
 \nonumber \\
\eea
where $g_{\parallel}, g_{\perp}>0$ are couplings in the $``z"$ and $``xy"$ directions in spin space.
As shown in Ref \cite{Japaridze}, in the limit of  large $g_{\parallel}$ and small $g_{\perp}$,  quantum fluctuations are small in the bulk. We shall therefore,  consistently with the classical  limit of the impurity spin, assume in the following that $g_{\parallel} \gg 1, g_{\perp} \ll 1$. 
Our approach in what follows is based on  the well-known correspondence between the XXZ deformation of the GN  and   sine-Gordon (SG) models. Upon bosonizing the fermions with
\be
\label{bosonization}
\psi^{'}_{L(R),a} = \kappa_a e^{-i \sqrt{\pi} (\theta_a \pm \phi_a)}/\sqrt{2\pi a_0}
\ee
where $\phi_{\uparrow,\downarrow}$ and $\theta_{\uparrow,\downarrow}$ are   boson fields and their duals and $\kappa_a$ are Klein factors satisfying $\{\kappa_a, \kappa_b\}= 2 \delta_{ab}$, the Hamiltonian (\ref{U1Thirring}) can  be  expressed as the sum of two commuting Hamiltonians describing charge  and spin fluctuations, i.e: $\mathcal{H}_{\text{bulk}} =
 \mathcal{H}_{c} + \mathcal{H}_{\rm SG}$, where 
\bea
\label{LL}
\mathcal{H}_{c}&=& \frac{1}{2}[ (\partial_x \phi_c)^2 + (\partial_x \theta_c)^2],  \\
\label{SG}
\mathcal{H}_{\rm SG}&=& \frac{1}{2}[ (\partial_x \phi)^2 + (\partial_x \theta)^2] -\frac{m_0^2}{\beta^2}
\cos(\beta \phi),
\eea
In the above equations (\ref{LL}) is the Luttinger liquid Hamiltonian which describes the 
massless charge degrees of freedom with $\phi_c=(\phi_{\uparrow} + \phi_{\downarrow})/\sqrt{2}$ and  $\theta_c=(\theta_{\uparrow} + \theta_{\downarrow})/\sqrt{2}$. The spin degrees of freedom of the fermions (\ref{bosonization})  are described  by the spin fields bosons, $\phi= 2\sqrt{\pi}(\phi_{\uparrow} - \phi_{\downarrow})/\beta$ and its dual $\theta=\beta(\theta_{\uparrow} - \theta_{\downarrow})/4\sqrt{\pi}$
whose dynamics is controlled by the  sine-Gordon (SG) Hamiltonian  (\ref{SG}). 
The parameters  $\beta$ and $m_0$ are related in a non-universal way to the couplings $g_{\parallel}$ and $g_{\perp}$ but in the limit we are interested with, which corresponds    to the semi-classical regime of (\ref{U1Thirring})  \cite{Japaridze},  we have $0<\beta^2 \le 4\pi$ and $m_0^2/\beta^2 \sim g_{\perp}/(\pi a_0)^2 \ll 1$ ($a_0$ being a short distance cut-off). 
The limit $\beta \rightarrow 0$ corresponds to the classical limit where the boson field is locked by the cosine potential
to one of its minima. When $\beta$ increases, quantum fluctuations become more important and are maximal
in this regime when $\beta=\sqrt{4 \pi}$.

In the bosonic  framework  the boundary twist on the fermions  (\ref{twistedOBC}) translates 
into  twisted boundary conditions on the  boson spin field $\phi$. Using (\ref{bosonization}) 
one finds 
\bea
\label{twistboson}
\phi(-L/2)&=&   \frac{2\pi}{\beta}n, \; \; 
  \phi(0)=\frac{4\alpha_0}{\beta}, 
 \eea
where  $n\in \mathbb{Z}$. Thus, given that
\be
 \label{spindensitybosonprime}
s^{'z}(x)= \frac{\beta}{4\pi}  \partial_x\phi(x),
\ee
the contribution of the boson to the spin of the system is 
 $s^{'z}= \alpha_0/\pi -n/2$. To get the total spin $s^z$ one must take into account
  the shift (\ref{transf}) which has a non trivial determinant. The easiest way to calculate its contribution 
to the spin density  is to use  bosonization and one eventually using Eqs.(\ref{S}, \ref{transf}, \ref{bosonization}) obtains
 \bea
\label{spindensityboson}
s^z(x) &=&-\frac{2\alpha_0}{\pi}\delta(x) + \frac{\beta}{4\pi}  \partial_x\phi(x).
\eea
Hence, with the twisted boundary conditions (\ref{twistboson}), the total  spin enclosed in the system
 is $\pm (1/2)$integer, as it should, i.e: $s^z=\int_{-L/2}^0 dx \; s^z(x) = -n/2$. 
 
 Before going further some comments are in order. As  is clear from (\ref{twistboson}),
 the boson dynamics is periodic  under the shift $\alpha_0 \rightarrow \alpha_0 + \pi/2$.
 The reason for this is that the boson field has radius $2\pi/\beta$. 
 In contrast,  the contribution coming from  the determinant of the  transformation (\ref{transf}) (the delta term in (\ref{spindensityboson})) corresponds to an accumulation of spin $-\alpha_0/\pi$ which is not periodic. One may however 
 restrict ourselves to the fundamental interval 
 \be\label{fundamental}
 \alpha_0 \in [0, \pi/2],
 \ee
keeping in mind that, due to the delta term in (\ref{spindensityboson}), 
the spin $s^z$ of any state is to be  shifted by $-n/2$ each time one shifts
$\alpha_0$ by $n \pi/2$.  One must be careful, however, when drawing any conclusions 
for  $\alpha_0 > \pi/2$ since the  contribution of the determinant  is 
 { \it localized} at the right edge of the system.   One should recall 
 that our { \it continuous} field theory  describes a system that cannot allow
 more than one fermion per spin  to be localized at the right edge. For instance if the 
  interaction between the impurity and the electrons is  local in real space,
  only one down spin fermion can be localized at the impurity site and $\alpha_0 \le \pi/2$.
  In this case  the local spin accumulation at the edge may  range between  $0$ and $-1/2$
  in the fundamental interval (\ref{fundamental}). This is the case that will be considered in this work.

\subsection{The classical limit of the bulk}
 
We  consider  the classical approximation   which corresponds to the limit $\beta \rightarrow 0$ in Eq.(\ref{SG}). In this  limit    the boson field $\phi(x)$ is locked in   one of the minima of the cosine potential in Eq.(\ref{SG}) located  at $\phi_p =   \frac{2\pi}{\beta}p$, $p\in \mathbb{Z}$.  The minimal energy configurations, consistent with the boundary conditions (\ref{twistboson}), correspond then  to   kinks or anti-kinks interpolating between one of the minima of the potential when $x=-L/2$ and $\phi=4\alpha_0/\beta$ at $x=0$. 
In the fundamental interval of twists, $0 \le \alpha_0 \le \pi/2$, one finds that the two minimal energy states consists of a kink $|\Psi_{+}\rangle$ interpolating between $\phi=0$ at $x=-L/2$ and  $\phi=4\alpha_0/\beta$ at $x=0$ and an anti-kink $|\Psi_{-}\rangle$ which interpolates between $\phi=2\pi/\beta$ at $x=-L/2$ and $\phi=4\alpha_0/\beta$ at $x=0$.  Their energies $E_{\pm}$  can be calculated using standard methods \cite{Rajaraman} and are given, up to a constant, by
 \bea
 \label{kinks}
 E_{+} &=&\frac{m_0}{\beta}\int_{0}^{4\alpha_0/\beta} d\phi\; \sqrt{2(1-\cos{\beta \phi}}), \; 
 \nonumber \\
 E_{-} &=&  \frac{ m_0}{\beta}\int_{4\alpha_0/\beta}^{2\pi/\beta} d\phi\; \sqrt{2(1-\cos{\beta \phi}}),
  \eea
  or
  \bea
  \label{energykink}
  E_{\pm}&=& \frac{1}{2}M (1  \mp\cos{2 \alpha_0}),
   \eea
where $M= 8m_0/\beta^2$ is the mass of the fundamental soliton of the SG
model.  The spins $s^{z}_{\pm}$ of the  kink and anti-kink  states are obtained from (\ref{spindensityboson}) 
and are given by
\be
\label{spinkink}
s^z_{+}=0, \; s^z_{-}=-\frac{1}{2}.
\ee
In the kink state the impurity remains unscreened with total spin (including that of the impurity)
$S^z_{\rm T}= S$ whereas in the anti-kink state it is partially screened with $S^z_{\rm T}= S-1/2$.
Depending on the value of the twist $\alpha_0$,  the ground-state consists of 
either the kink or the anti-kink state. From (\ref{energykink}) one can easily find that the ground-state
is given by 
\bea
\label{gspm}
 |\Psi_{+}\rangle\;  {\rm when}\; 0 \le \alpha_0 \le  \frac{\pi}{4}, \; \; \; 
 |\Psi_{-}\rangle \;  {\rm when}\; \frac{\pi}{4} \le \alpha_0 \le \frac{\pi}{2}
\eea
and become degenerate at $\alpha_0= \pi/4$ when the two states cross.
In each  of the domains (\ref{gspm}), the first excited state is given by   the anti-kink $|\Psi_{-}\rangle$ or the kink $|\Psi_{+}\rangle$ respectively and  has an energy gap
\begin{center}
\begin{figure}[!h]
\includegraphics[width=0.7\columnwidth]{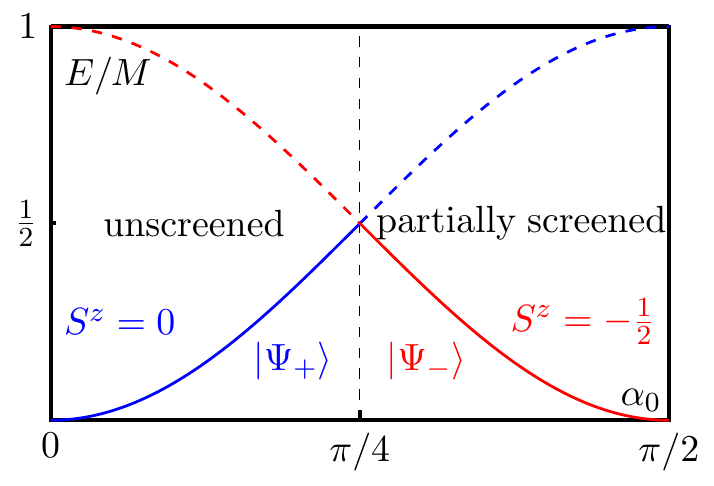}
\caption{Kink and anti-kink boundary states and excitations with energies given as a function of the twist angle $\alpha_0=JS/2$.  $S^z$
denotes the spin of the bulk fermions which is zero in the unscreened phase and $-1/2$ in the 
partially screened phase. Both the kink and the anti-kink states are degenerated at  the quantum
 (boundary) phase transition point at  $\alpha_0=\pi/4$.}\label{fig1}
\end{figure}
\end{center}

\be
\label{midgapsemiclassical}
\epsilon= |E_{+} -E_{-}|=M | \cos{2 \alpha_0}|,
\ee
above the ground-state.  As $\epsilon \le M$, these states lie in the superconducting
gap below  the  continuum made of propagating solitons or anti-solitons. These mid-gap
states are  the one dimensional analogue of the YSR states found in the mean-field approach.
The  phase diagram in the fundamental period $\alpha_0 \in[0, \pi/2]$ is depicted in  FIG. \ref{fig1}.
In the regime $\pi/4 \le \alpha_0 \le \pi/2$ the impurity spin is partially screened in the ground-state which has total spin  $S^z_{\rm T}=S_{}-1/2$. In the domain $0 \le \alpha_0 \le \pi/4$ the total spin is $S^z_{\rm T}=S_{\rm}$ and the impurity spin remains unscreened. At $\alpha_0 = \pi/4$ a first order quantum phase transition occurs, between unscreened and a partially screened phase, where the mid-gap (\ref{midgapsemiclassical}) closes.  

One may finally observe that the phase diagram  
 qualitatively matches with that obtained from the mean-field or BCS approach in the range of twists
 in the fundamental interval $\alpha_0\in [0, \pi/2]$.
 Both predict a phase transition between unscreened and screened phases driven by an YSR level
 but with different critical values of the twist or exchange interaction, i.e: $\alpha_0=\pi/4$ ($(JS)_c= 2/\pi N_0$) and $\alpha_0=1$ ($(JS)_c= 1/2 N_0$)  for the semi-classical and BCS   case respectively. 
 

 \subsection{Quantum bulk at the Luther-Emery point}
 
 We would like now to investigate the effects of the quantum fluctuations in the bulk, i.e. the effect of a non-zero
 $\beta$ in (\ref{SG}). One route would be to calculate perturbatively the fluctuations around the semi-classical kink and anti-kink
 configurations  $|\Psi_{\pm}\rangle$ defined above. For the purpose of the present work we shall rather content ourselves
 by solving the model at the value of  $\beta^2= 4\pi$. For this value of the interaction,
 the SG model can be mapped onto a model  of free  massive spinless fermions and can be solved exactly. 
 This is the so-called Luther-Emery point~\cite{LutherEmery}. To proceed further we  define  the left and right moving Luther-Emery fermions 
 \be 
 \label{LE}
 \Psi_{L(R)}= e^{\mp i\sqrt{4\pi}\phi_{L(R)}}/\sqrt{2\pi a_0},
 \ee
 where $\phi_{L(R)}=(\phi \pm \theta)/2$ are left and right moving boson operators and 
 $a_0$ is a short distance cut-off. With help of (\ref{LE}) one can
  rewrite the SG Hamiltonian (\ref{SG}) as
 \be
 \label{LEH}
 H_{}= \int_{-L/2}^{0} dx\; \Psi^{\dagger}(x) h_{D}   \Psi^{}(x),
 \ee
 where $\Psi^{T}=( \Psi_{R}, \Psi_{L})$ is a two-components spinor and 
 \be
 \label{dirac}
 h_D=-(i\sigma^z \partial_x + m \sigma^y),
 \ee
 is the Dirac Hamiltonian where $m=g_{\perp}/(\pi a_0) > 0$ is a mass parameter.
 The spin density operator (\ref{spindensityboson}) expressed in terms of the Luther-Emery
 fermions is 
 \be
 \label{spindensityLE}
 s^z(x)= -\frac{2 \alpha_0}{\pi} \delta(x) + \frac{1}{2} \Psi^{\dagger}(x) \Psi^{}(x).
 \ee
 To complete the mapping we also need to translate the boundary conditions of the boson field (\ref{twistboson}) 
 into boundary conditions of the spinor field (\ref{LE})
\be
\label{BCLEF}
\Psi_R(-L/2)=\Psi_L(-L/2), \; \Psi_R(0)=- e^{i 4\alpha_0}\Psi_L(0),
\ee
which are obviously invariant under the shift 
$\alpha_0 \rightarrow \alpha_0 + \pi/2$ as expected. With these results at hand,
 the physics at the Luther-Emery point boils down to the 
  knowledge of the solutions of  the Dirac equation
 $
h_D\Psi(x,t) = i\partial_t\Psi(x,t)
$
with the boundary conditions (\ref{BCLEF}). As we shall focus on the ground-state
properties we shall need only the negative energy solutions. 

The  wave-functions corresponding to   negative energies $E_{\lambda} =-m\cosh{(\pi \lambda)}$ and pseudo-momenta $p_{\lambda}  =m \sinh{(\pi \lambda)}$ are given by $ \chi_{\lambda}(x,t) =  e^{-iE_{\lambda} t}\;   \chi_{\lambda}(x)$ where
\bea
 \label{soldirac-}
 \chi_{\lambda}(x) = 
 a_{\lambda}  \begin{pmatrix}- i e^{-\pi \lambda/2} \\  e^{\pi \lambda/2}\end{pmatrix} e^{ip_{\lambda} x}+
   a_{-\lambda}  \begin{pmatrix} -i e^{\pi \lambda/2} \\  e^{-\pi \lambda/2}\end{pmatrix} e^{-ip_{\lambda} x}. \nonumber \\
  \eea
 The amplitudes $a_{\lambda}$ and $a_{-\lambda}$ are obtained  (up to a normalization) by matching the right boundary condition at $x=0$ (\ref{BCLEF}) which  gives the impurity S-matrix  
 \be
 \label{SLE-}
 S^-_{\rm imp}(\lambda)= a_{\lambda} /a_{-\lambda} = - \frac{\cosh{\frac{\pi}{2}(\lambda- i (\frac{1}{2}-a))}}{\cosh{\frac{\pi}{2}(\lambda+ i (\frac{1}{2}-a))}}.
 \ee
 The boundary condition at the left boundary $x=-L/2$  itself yields to  the Bethe equation
 \bea
 \label{BELE-}
e^{im \sinh{(\pi \lambda)}L}= \frac{\cosh{\frac{\pi}{2}(\lambda-i(\frac{1}{2} -a))} \cosh{\frac{\pi}{2}(\lambda-\frac{i}{2})} }{\cosh{\frac{\pi}{2}(\lambda+i(\frac{1}{2} -a))} \cosh{\frac{\pi}{2}(\lambda+\frac{i}{2})}}.
 \eea
In the above equations, we have introduced the parameter
 \be
 \label{aLE}
 a=2-\frac{4\alpha_0}{\pi},
\ee
instead of $\alpha_0$ for convenience. In terms of $a$, the periodicity of the boundary conditions, i.e: 
$\alpha_0 \rightarrow \alpha_0 + \pi/2$, translates into $a \rightarrow a-2$ and the fundamental period
$\alpha_0 \in [0, \pi/2]$ is given by $a \in [0,2]$.

Boundary bound-states are obtained by looking at the zeros or the poles of the  S-matrix (\ref{SLE-}) which are solutions of the Bethe equations. Both  eventually lead to the same  bound-state   localized at the right boundary close to $x=0$ corresponding to the  imaginary roots
 $ \lambda=\pm i(1/2+a)
 $.
They both give  the same bound-state solution  with energy
 \be
 \label{BSELE}
 \epsilon=m \sin{(\pi a)},
 \ee
 and wave-function
 \be
 \label{BSWF}
 \chi_{\rm b.s}(x) = 
 \xi_a^{-1/2} \begin{pmatrix} -e^{-i\pi a/2} \\  e^{i\pi a/2}\end{pmatrix} e^{ x/\xi_a},
  \ee
 where 
 \be\label{xia}
 \xi_a = (-m\cos{(\pi a)})^{-1}.
 \ee
 The bound-state wave-function being normalizable only when $m\cos{(\pi a)} < 0$,  the bound-state exists only in the mid-gap domain
 \be
 \label{midgapinterval}
   a \in [ \frac{1}{2},  \frac{3}{2}] \Leftrightarrow \alpha_0 \in [\frac{\pi}{8}, \frac{3\pi}{8}], 
       \ee
and is localized  close to  the impurity  site with a characteristic length scale  $\xi_a$. 
The above bound-state solution corresponds to the mid-gap state found in the semi-classical analysis
and in previous mean-field approaches.
However, in sharp contrast with the former analysis, a mid-gap state exists  only in a narrowed domain of twists $\alpha_0$, or $a$, at the Luther-Emery point.

\subsubsection{Ground-state}

The ground-state of the system is obtained by "filling" the vacuum  with all negative energy states. 
In the region $a\in[1/2, 3/2]$ it  may or may not include the bound-state (\ref{BSWF}) depending on whether 
 $\epsilon$ is negative or positive. In the following we shall distinguish between  the fermion operator $b^{\dagger}$  that creates   the bound-state (\ref{BSWF})  and $(a^{-}_{\lambda})^{\dagger}$, with $\lambda$ real, those associated with the propagating negative energy modes $\chi^*_{\lambda}(x)$. We shall accordingly  consider the two states
\be
\label{FS}
|\Omega \rangle= \Pi_{\lambda}^{} (a^{-}_{\lambda})^{\dagger} |0\rangle,\;\; \;  |B\rangle \equiv b^{\dagger} |\Omega \rangle,
\ee
and compare their energies as a function of $a$.  After a straightforward calculation one finds that
the ground-state $|GS \rangle $ is eventually  given   by
\bea
\label{GS}
|GS \rangle &\equiv& |\Omega \rangle, \; \; \; \; a\in  \; [0, 1] \; {\rm and}\;  a\in  \; [ \frac{3}{2}, 2 ],\nonumber\\
|GS \rangle &\equiv& |B\rangle,  \;  \;  \;  \;   a\in  \;[1, \frac{3}{2}],
\eea
with energy $E_{\rm GS}(a)$ given, up to a constant,  by:
\bea
\label{EGS}
E_{\rm GS}(a)&=&E_{\Omega}(a), \; \; \; \; \; \; \; \; \; \; \; \;  \; \; \; \;  \; \; \; \;  a\in  \; [0, 1] \; {\rm and}\;  a\in  \; [ \frac{3}{2}, 2 ],\nonumber\\
E_{\rm GS}(a)&=&E_{\Omega}(a) + m \sin{\pi a} \; \;  \; \;  a\in  \;[1, \frac{3}{2}],
\eea
where 
$E_{\Omega}(a)= m + m \sin{\pi a}\;  \tan^{-1}{[\tan{( \pi/4 -\pi a/2)}]}/\pi$ when  $a\neq  3/2$
 and  $E_{\Omega}(a)=m/2$ when $a=  \frac{3}{2}$ is the energy of the Fermi sea.
In the intervals $a \in [1/2, 1 ]$ and $a \in [ 1, 3/2]$
the states $|B \rangle$ and $ |\Omega \rangle$ are 
mid-gap states with an energy $\pm m \sin{\pi a}$ above the ground-state respectively. 
The two states cross at  $a=1$, or $\alpha_0=\pi/4$, where the ground-state is doubly
degenerated and a first order phase transition occurs.  

The spin quantum numbers of the states $|\Omega \rangle$ and $ |B\rangle$
can be obtained from Eq.(\ref{spindensityLE}). Denoting by 
 $s^z_{\Omega}(a)$ the spin of the Fermi sea, since    the state  $|B>=b^{\dagger}|\Omega \rangle$ contains one more Luther-Emery fermion, its spin  is given by $s^z_B(a)=s^z_{\Omega}(a) + 1/2$. The calculation is standard and one finds
 for $s^z_{\rm GS}(a)=\langle {\rm GS}| s^z|{\rm GS}\rangle$  in the fundamental interval $a\in[0,2]$
\bea
\label{SGS}
s^z_{\rm GS}(a)=   -\frac{1}{2},  \; a\in  \; [0, 1]; \; S^z_{\rm GS}(a)= 0,\; \; a\in  \; [ 1, 2].
\eea
In the mid-gap region (\ref{midgapinterval}) the two mid-gap states $ |B \rangle$ and $ |\Omega \rangle$  in the  sectors $a \in [1/2, 1] $ and $a \in [ 1, \frac{3}{2}]$ carry an additional spin $\Delta S^z=\pm \frac{1}{2}$  with respect to that of the ground-state respectively .

 When comparing to the semi-classical phase diagram of 
(\ref{fig1}) we see that both qualitatively match in the mid-gap interval $a \in [1/2, 3/2] \Leftrightarrow \alpha_0 \in [\pi/8, 3\pi/8]$. In particular  we observe that the two analyses predict a  quantum critical point  between the unscreened and a partially screened phase 
at the same point $a=1$ or $\alpha_0=\pi/4$. However for larger departures from the phase transition point they differ significantly in that  the mid-gap region,
where a bound-state exists,  gets {\it narrowed} to  the interval $a \in [1/2,3/2]$  (i.e: $\alpha_0 \in [3\pi/8, \pi/8])$ at the Luther-Emery point. 
This is due to the quantum fluctuations in the bulk. 
As a consequence, both the partially screened phases ($a<1)$ and  the unscreened phase ($a>1$) are split into two regimes which depend  on the existence,  or not, of a bound-state in the mid-gap region. 
\begin{center}
\begin{figure}[!h]
\includegraphics[width=0.7\columnwidth]{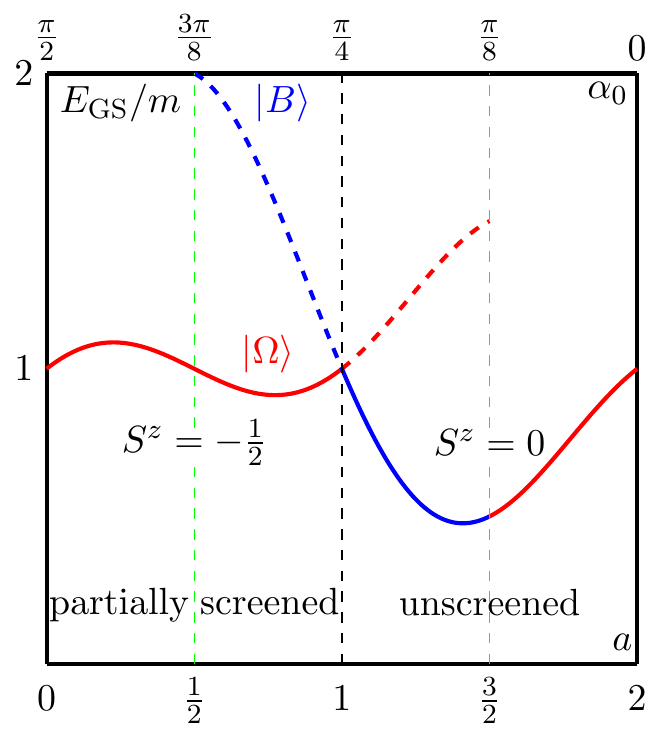}
\caption{Phase diagram in the fundamental interval $a\in[0,2] \Leftrightarrow \alpha_0 \in [0, \pi/2]$ at the Luther-Emery point.
The  curve in red denotes the energy of the Fermi sea $|\Omega \rangle$ while the one in blue that of the state $|B\rangle$
where the bound-state is present. The ground-state is represented by a solid line whereas the mid-gap YSR state by a dashed line.
The vertical dashed green lines define the domain where the YSR state exists.}\label{GSEnergy}
\end{figure}
\end{center}

We display the phase diagram in the fundamental interval $a\in[0,2] \Leftrightarrow \alpha_0 \in [0, \pi/2]$ in the Fig.(\ref{GSEnergy}).
The observations made above lead us to the following picture for the screening process. 
When $\alpha_0=0$, the impurity decouples form the superconductor which is in a singlet state
with $S^z=0$. As  one increases the coupling to the impurity $J$, $\alpha_0$ also increases and 
nothing dramatic happens until it reaches the value $\pi/8$. In the region $0 \le \alpha_0 < \pi/8$
the ground-state is a singlet with $S^z=0$. Above the gap, at energies $E\ge m$, the excitations consist into spin $S^z =\pm 1/2$ propagating electrons or holes. When $\alpha_0 \ge \pi/8$ a hole
state, with spin $S^z =- 1/2$, is dragged from the continuum  into the mid-gap region where it becomes a bound-state  localized close to  the boundary with a localization length $\xi_a$ (\ref{xia}). When $\alpha_0\sim \pi/8$, $\xi_a$ is large and the trapped spin $-1/2$ particle is almost delocalized in the bulk with a vanishingly small amplitude. However when  $\alpha_0 > \pi/8$  the particle is getting more and more localized with $\xi_a$  reaching its minimum value $\xi=1/2m$ when $\alpha_0= \pi/4$ at the critical point. At this point this state become the ground-state and   has a total spin $S^z=-1/2$ thus
partially screening the impurity. As $\alpha_0$ further increases this state
  remains the ground-state until one reaches  $\alpha_0=\pi/2$.

In this section we have solved the model  with a classical impurity spin in the two limits  $\beta \rightarrow 0$ and $\beta =\sqrt{4 \pi}$.  In both limits a two phase structure emerges, corresponding to an unscreened and a partially screened impurity,  separated by  a quantum critical point at $\alpha_0=\pi/4$ which is unaffected by quantum fluctuations. 
Our solutions demonstrate that the main effect of quantum fluctuation is to reduce, in the fundamental interval, by half 
the domain in which YSR states exist. Despite this, in both cases, the quantum phase transition is still driven by the intra-gap YSR state. We now turn to study the effect of the dynamics of the impurity spin to inquire how, as in the BCS case, this scheme
is modified by the  quantum fluctuations of the impurity degrees of freedom generated by spin flip processes.

\section{ Quantum impurity: The Bethe Ansatz results\label{bethe}}

In this section we discuss the results of  the solution of the model obtained by the Bethe Ansatz, relegating the details of the calculation to the appendix. This method allows access to the exact spectrum and has been used recently to investigate boundary and impurity effects in interacting field theories~\cite{RylandsAndrei1,RylandsAndrei2, RylandsAndrei3, PRA,  Rylands, PAN1, PAN2}. In particular, it was shown in Ref.[\citenum{PRA}] that  the Hamiltonian (\ref{Hamiltonian}) is integrable for generic values 
of the bulk coupling $g$ and contact exchange interaction $J$. 
The resulting physics is described in terms of two Renormalization Group (RG) invariants:
the superconducting gap $\Delta$ and a parameter $d$ that labels the RG invariant flow lines
in the $(g, J)$ plane. In terms of the latter couplings, $\Delta$ and $d$ express as
\bea
\Delta&=&D\arctan{(\frac{1}{\sinh{\pi b}})} \\
d&=&\sqrt{b^2-2b/c -1},\label{RGinv}
\eea
where $D$ is a cut-off scale and
\bea \label{couplings}
b=(1-g^2/4)/2g,\; c=2J/(1-3J^2/4).
\eea
The RG invariant $d$ may be real or purely imaginary in which case we shall use the notation $d=ia$. 
Notice that although we are using the same symbol $a$ as for the twist in the previous section,
here $a$ depends on {\it both} $J$ and $g$.
Preliminary investigations  of the Bethe equations  associated with (\ref{Hamiltonian}) have revealed    
 a two phases structure:  a screened Kondo  phase when $d \gg 1$ corresponding to $J\gg g$  and an unscreened phase when $a \gg 1$ corresponding to $g \gg J$. 
 In the following we extend the analysis made in  \cite{PRA} to the full domain 
of couplings $d$ and $a$. We shall distinguish between three regimes of $d$ or $a$
where the Bethe equations take different forms and have different types of solutions. 
First is the regime of $a \ge 3/2$ that we shall label the unscreened regime since in this region the impurity
 remains always  unscreened. The second region of couplings is obtained when   $d$ is real and $a \le 1/2$. 
 It is  a Kondo-screened regime where the impurity is always screened in a Kondo-like fashion. 
 Finally we shall call  the third regime, i.e: $1/2 < a < 3/2$,  the YSR regime. In the scaling limit,as 
 $\Delta$ and $d$ are held fixed while $D\to \infty$ and both $J,g \to 0$ it corresponds to a narrow  sliver in coupling space. In this regime 
 a localized bound-state appears with energy  within the superconducting gap. As in the BCS case screened and unscreened
 states are obtained  by adding or removing the bound-state and crossing at $a=1$
 where a boundary quantum phase transition occurs between a screened phase when $a<1$ to an
 unscreened phase when $a >1$.

\subsection{The unscreened regime: $a \ge 3/2$.\label{unscreenedregime} }

This phase obtains in the purely imaginary $d=ia$ regime
when $a\ge 3/2$.  The ground-state  is  doubly degenerated, 
\be\label{doublet}
|{\rm GS}\rangle \equiv \{ |+ 1/2\rangle, |- 1/2\rangle\},
\ee
having  a total spin $ S_{\rm T}^z = \pm 1/2$ and  an even fermion parity  ${\cal P}=(-1)^{N}=+1$.
In this phase   the impurity spin remains unscreened. Holon excitations, which are  charged spin-singlet excitations, are gapless
while in the spin sector all excitations have energies   above the  gap $\Delta$. In contrast to the BCS model there are no intra-gap YSR
bound-states.

\subsection{The Kondo-screened regime: $ a \le 1/2$ and $d \in \mathbb{R}$.} 

For $a \le 1/2$  or when $d$ is real, a Kondo-screened regime is obtained.
The ground-state  of  (\ref{Hamiltonian}) is a spin singlet, 
\be\label{singlet}
|{\rm GS}\rangle \equiv |0\rangle,
\ee
with total spin $\vec S_{\rm T}=0$  and  an odd fermion parity, i.e: ${\cal P}=(-1)^{N}=-1$.  
As in the unscreened phase, on top of the gapless holon excitations, all excitations are above
the superconducting gap. There are no intra-gap states in sharp contrast with results in
 mean-field or BCS case. In particular one cannot interpret the  singlet state $|0\rangle$ (\ref{singlet}) in the screened regime as  a mid-gap state obtained by adding a localized bound-state close to  the right edge in one of the doublet ground-states  $|\pm 1/2 \rangle$ (\ref{doublet}) of the unscreend regime. As we shall argue, the screening of the impurity resembles a genuine many body Kondo effect in a metallic phase.

The Kondo-screened regime is characterized  by the generation of an 
energy scale $T_K$ which is the analogue of the Kondo temperature 
in the metallic phase. It must be expressed in terms of the RG invariants, $T_K= \Delta f(d)$ with the choice of $f(d)$ depending on the particular question examined. Here we determine it by considering  the ratio of impurity density of states (DOS)
to that of the bulk DOS per unit length as given in Ref.[\citenum{PRA}]
\be\label{DOS}
 R(E)= \frac{L}{2} \frac{ \rho_{\rm imp}(E)}{ \rho_{\rm bulk}(E)}= \frac{\Delta \cosh{\pi d}}{(E^2+\Delta^2 \sinh^2{\pi d})}, \; E \ge \Delta,
\ee
where $\displaystyle{ \rho_{\rm bulk}(E)=\frac{E}{\pi \sqrt{E^2-\Delta^2}}}$,
and  define 
\be\label{TK}
T_K= \Delta \sqrt{1+ \cosh^2{\pi d}},
\ee
 as   the width of $R(E)$ at half its maximum. In the Kondo-screened regime $T_K \ge \Delta$ 
 for all $d$ real and $d=ia$. It eventually  reaches
 its minimum value at $a=1/2$ where $T_K = \Delta$.
 
 In the limit of large real $d \gg 1$,  $T_K/\Delta \gg 1$ and  $T_K$ defines an energy scale which is sharply different from the superconducting gap. In this regime  (\ref{DOS}) takes a characteristic Loretzian form  $R(E)  \simeq T_K/(E^2 + T_K^2)$ with $T_K\simeq \frac{\Delta}{2} e^{\pi d} \gg \Delta$. We then deduce that the   impurity density of state in the large $E$ limit
 \be\label{DOSKondo}
\rho_{\rm imp}(E) \simeq \frac{T_K}{\pi (E^2 +T_K^2)}, \; E \gg \Delta,
\ee
matches with that of the Kondo model in the normal state with a  Kondo temperature $T_K$. In particular, 
 in the massles limit $(g, \Delta) \rightarrow  0$,   $T_K \rightarrow T_K^0$ and $\rho_{\rm imp}(E)$
matches with impurity DOS in the normal phase. 
In this respect, the scale $T_K$, as given by (\ref{TK}), may be seen as 
a {\it renormalized} Kondo temperature which depends on both $J$ and $g$ through the gap $\Delta$ and the RG invariant $d$, as noted above.
In the regime $T_K/\Delta \gg 1$ the scale $T_K$ distinguishes, exactly as in the metallic phase, between the strong and weak coupling regimes of the Kondo interaction $J$. Indeed, using the RG equations obtained in [\citenum{PRA}], one finds  that at some scale $\Delta \ll \Lambda \le D$ the coupling $(c,b)$ scale as
 \be
 c_{\Lambda}= \frac{2 b_{\Lambda}}{ b_{\Lambda}^2-(1+d^2)}, \; b_{\Lambda} \simeq \frac{1}{\pi} \ln{\frac{2\Lambda}{\Delta}} \gg 1.
 \ee
 We immediately see that   $c_{\Lambda}$ diverges when  $\Lambda \rightarrow T_K$, indicating that $J$ reaches a strong coupling regime while, since $b_{T_K} \simeq d \gg 1$,   the bulk coupling $g$ 
 remains small.   Hence, despite the presence of the superconducting gap $\Delta$, the condition $T_K \gg \Delta$ insures that $J$ 
reaches the strong coupling regime while the bulk coupling $g$ is still in the scaling region. 
Therefore,in the deep Kondo regime $d\gg 1$, the screening of the impurity in this region is similar of the Kondo screening in the metallic phase.

For other values of the couplings, i.e:  $d\simeq 1$ or $d=ia$ with $a < 1/2$, we have
$T_K/\Delta \simeq 1$ and both $J$ and $g$ reach the strong coupling regime 
when $\Lambda \simeq T_K$. Although the screening of the impurity competes with the formation of Cooper pairs,
the impurity eventually gets screened as our exact solution shows. 
The main difference between  the large $T_K$ and  $T_K \simeq \Delta$ regimes 
lies in the fact that the DOS ratio $R(E)$ 
gets more and more peaked close to $E = \Delta$ as $T_K$  approaches $\Delta$.
In particular we notice  that when $a \rightarrow 1/2$, $R(E)\rightarrow 0$ for all values of $E \neq \Delta$.
We interpret this fact as a cross-over from many-body screening to  single particle screening as  $T_K$
range from $T_K \gg \Delta$ to $T_K = \Delta$. As we shall now see, when $1/2 < a < 3/2$, a different situation emerges
in which the screening of the impurity is related to a localized   bound-state mode in the mid-gap region.

\subsection{The YSR regime: $1/2 \le a \le 3/2$.} 

In this regime of couplings the states with the  lowest energy may have different fermion parities ${\cal P} = \pm 1$.
In the even fermion parity sector, i.e: ${\cal P} = +1$, there exists two degenerate states, $ |\pm 1/2\rangle$,
with total spins $ S^z_{\rm T}=\pm 1/2$ as in the unscreened regime (\ref{unscreenedregime}). On top of that,
there exists also a state $ |0'\rangle$ with  total spin  $\vec S_{\rm T}=0$ and  odd fermion parity ${\cal P} = -1$.
This state is obtained from the latter by adding a {\it bound-state}  localized at the right boundary $x=0$ 
to either of the states $ |+ 1/2\rangle$ or $|- 1/2\rangle$.
In the Bethe Ansatz framework, such a bound-state  corresponds  to a  boundary string which arises as a purely imaginary solutions of the Bethe equations. 

The energy of the bound-state is found to be the energy difference between the ground-states  with total spins $ S^z_{\rm T}=0$ and $ S^z_{\rm T}=\pm 1/2$, 
\be
E_B= E^0_N- \frac{1}{2}(E^0_{N-1}+E^0_{N+1}),
\ee
where $E^0_N$ is the ground-state energy of the system with an  odd number of fermions $N$
and  total spin $S^z_{\rm T}=0$ 
and $E^0_{N\pm 1}$ are the energies of the states with an even number of particles and total spins $ S^z_{\rm T}=\pm 1/2$.
The energy $E_B$ precisely measures the energy cost, up to the chemical potential,  of adding an electron to the system.
It consists of a charge part and a spin part, i.e:  
$
E_B=E_{\rm charge}+ \epsilon,
$
where, 
\be\label{Echarge}
E_{\rm charge} = -\frac{\pi}{2L},
\ee
and the spin part
\be\label{boundstateenergy}
\epsilon = -\Delta \sin{\pi a},
\ee
is the energy of the bound-state in the thermodynamic
limit. Since it is  always {\it smaller} that the superconducting gap  
 the ground state of the system is the given, in the limit $L \rightarrow \infty$, by 
\bea
|{\rm GS}\rangle &\equiv & |0'\rangle, \;  \; \; \; \; \; \; \frac{1}{2} \le a \le 1, \nonumber \\
|{\rm GS}\rangle &\equiv  & |\pm 1/2\rangle, \; 1 \le a \le \frac{3}{2}.
\eea
In the regime $1 < a < 3/2$, the ground-state is doubly degenerate and the impurity is unscreened.
Adding a bound-state to the system produces the singlet state $|0'\rangle$ in which the impurity is screened. 
This state has an energy  $|\epsilon| < \Delta$ above the ground-state and is hence a mid-gap state.
In the regime $1/2< a < 1$, the screened state and unscreened  states switch their roles. The ground-state
 is given by $|0'\rangle$ and the impurity is screened while the two degenerate unscreened states
$ |\pm 1/2\rangle$ are mid-gap states at an energy $|\epsilon|$ above the ground-state. Finally at 
$a=1$, $\epsilon=0$, the three states $ |\pm 1/2\rangle$ and  $|0'\rangle$ are degenerate and there
is a first order quantum phase transition between screened and unscreened phases. We notice that, thanks 
to (\ref{Echarge}), adding or removing an electron when $a=1$ costs no energy in the thermodynamic limit.
This corresponds to a genuine zero-energy mode in the system. The situation at hand is similar to what happens
in the BCS case.
 
\section{Discussion}\label{discussions}
In this paper we have studied the physics of a magnetic impurity coupled to a superconductor with a view to understanding the role of quantum fluctuations. We considered a charge conserving $s$-wave superconductor which exhibits several regimes depending on the strength of the fluctuations. We have established that bulk quantum fluctuations can drastically affect the phase diagram inferred using mean-field methods for the BCS model.  

We first considered a classical impurity system in the limit where bulk quantum fluctuations are minimal. In this case, using bosonization, it was seen that the YSR state exists throughout the phase diagram. Subsequently, the quantum fluctuations were reintroduced by studying the classical impurity system at the Luther Emery point via refermionization. It was shown that the YSR state is destabilized in half of the phase diagram. 

Finally, the full problem of a quantum impurity and quantum fluctuating $SU(2)$-invariant bulk was studied using Bethe Ansatz revealing a rich structure with three regimes named Kondo, YSR and unscreened. 
The main effect of the quantum fluctuations is the suppression of the YSR state everywhere except in the narrow YSR regime around the quantum critical point. Further, the quantum nature of the impurity reveals itself  in the Kondo regime where  a renormalized Kondo effect takes place.  This is  due to the strong quantum fluctuations resulting from high scattering rate between the electrons in the bulk and the impurity, there exists no YSR states. However when the impurity coupling strength is decreased such that $T_K\simeq \Delta$, a smooth crossover from the many-body screening to a single particle screening of the impurity takes place, and the system enters YSR regime. In this regime, the effect of the bulk fluctuations on the quantum impurity is drastically reduced and the  YSR states appear. 

As  in the BCS case, we expect that
  this regime of the phase diagram can be  described by treating the impurity as classical
  provided one uses a {\it renormalized} value of the twist $a$ or $\alpha_0$ 
  which depends on both $g$ and $J$. This was explicitly  shown to be the case at the special anisotropic (XXZ bulk)  Luther-Emery point  where we found that 
  the renormalized value of $a$  in Eq.(\ref{BSELE}) is given
  by Eqs.(\ref{RGinv}, \ref{couplings}). Preliminary calculations made for an $SU(2)$ invariant
  bulk, as well as for small departures from the Luther-Emery point, show that this feature also holds in the more general case. The domain of stability of the YSR state narrows as one increases the strength of the quantum fluctuations. These results will be
 published in a separate paper \cite{private}. In any case, in the YSR regime the impurity is screened by a local single particle spin accumulation, and a level crossing occurs between the screened and unscreened states which correspond to presence and absence of the spin accumulation. 

Our results highlight the prominent role that quantum
fluctuations can play in the physics of YSR states. Although  they have been derived in the specific  example  of  a  one-dimensional charge  conserving superconductor they suggest that YSR states in other non-conventional superconductors, where fluctuations  are  prominent,  including  systems in higher  dimensions such as d-wave superconductors,  need to be reexamined. Furthermore, we have studied the case of a single impurity however, applications of YSR states to topological superconductors require chains of such impurities and the hybridization of many YSR states. The effects of quantum fluctuations on such multiple impurity systems shall be the focus of a forthcoming paper. 

\acknowledgements{}
PRP is supported by Rutgers university HEERF fellowship. CR acknowledges support from ERC under Consolidator grant  number 771536 (NEMO)

\bibliography{refks}

\appendix
\begin{widetext}
\section{Bethe Ansatz Solution}
\label{Bethe}

The Bethe equations were derived in \cite{PRA}. In section \ref{BetheequationsAp} we present a brief overview of the derivation of the Bethe equations. In section \ref{BethesolutionAp} we present the solution to the Bethe equations in all the regimes.

\subsection{Bethe Equations}
\label{BetheequationsAp}
 Since the  Hamiltonian (\ref{Hamiltonian}) commutes with total particle number $ N =  \int dx\;  \left( \psi^{\dagger}_{L}\psi_{L} +  \psi^{\dagger}_{R}\psi_{R}\right)$, $\mathcal{H}$ can be diagonalized by constructing the exact eigenstates in each $N$ sector. From here on, for notational convenience,  we shall use the notation  $(+, -)$  to indicate the chirality index of the fermions replacing $(R, L)$ notation. The $N$-particle eigenstate takes the standard reflection Bethe Ansatz form of a plane wave expansion in  $N!\, 2^N$ different regions of coordinate space. The state is labeled by momenta $k_j, j=1\cdots N$, the same in all regions,  and  is given by,
\bea\nonumber
\ket{\{k_j\}}=
\sum_{\substack{\{a_j\},\{\sigma_j\}\\Q}}\int \mathrm{d}\vec{x}\,F^{\{\sigma\}}_{\{a\}}(\vec{x})\prod_{j=1}^N \psi^{\dagger}_{\sigma_j,a_j}(x_j)\ket{0}
\otimes \ket{a_0}.
\end{eqnarray}
where $F^{\{\sigma\}}_{\{a\}}(\vec{x})=\theta(x_Q) A^{\{\sigma\}}_{\{a\}}[Q] e^{i\sum_{j}^N\sigma_j k_jx_j}$ is the $N$-particle wavefunction. 
Here the sum is over all different chirality and spin configurations specified by $\{\sigma_j\}=\{\sigma_1,\cdots ,\sigma_N\}$, $\{a_j\}=\{a_0,a_1,\cdots a_N\}$ where $\sigma_j=\pm$ and $a_j=\uparrow, \downarrow$ are the chirality and spin of the $j^\text{th}$ particle. $a_0$ denotes the spin of the impurity with $\ket{a_0}$ being the state of the impurity and $\ket{0}$ is the vacuum which contains no particles. We also sum over all different orderings of the particles, labelled by $Q$ which are elements of the symmetric group $Q\in \mathcal{S}_N$. $A^{\{\sigma\}}_{\{a\}}[Q]$ are the amplitudes for a particular spin and chirality configuration and  ordering of the particles while $\theta(x_Q)$ is a Heaviside function which is non zero only for the ordering of particles labelled by $Q$. These amplitudes are related to each other via application of the various $S$-matrices in the model which are determined by the $N$-particle Schr\"{o}dinger equation and the consistency of the solution. Amplitudes which differ by changing the chirality of the rightmost particle $+ \rightarrow -$ are related by the particle-impurity S-matrix $S^{j0}$ and the amplitudes which are related by swapping the order of particles with different chiralities are related by the particle-particle S-matrix $S^{ij}$. These are given by \begin{eqnarray}\label{S12}S^{ij}= \frac{2ib \; I^{ij} +P^{ij}}{2ib+1},~
S^{j0}=\frac{\; I^{j0} -i cP^{j0}}{1-ic}\end{eqnarray}
where $I^{ij}$ is the identity operator
 and $P^{ij}=(I^{ij}+ \vec{\sigma}_i\cdot\vec{\sigma}_j)/2$ is the permutation operator acting on the spin spaces of particles $i$ and $j$ with $0$ indicating the impurity. Furthermore we have introduced the parameters $b= \frac{1-3g^2/4}{4g}$ and  $c=\frac{2J}{1-3J^2/4}$ which encode separately the bulk and impurity coupling constants. An additional  $S$-matrix, denoted by $W^{ij}$, is also required. It  relates amplitudes that differ by exchanging particles of the same chirality. This is given by $W^{ij}=P^{ij}$. The consistency of the solution is then guaranteed as the $S$-matrices satisfy the Yang-Baxter and Reflection equations
 \cite{Sklyannin, Cherednik}.

Imposing the boundary condition at $x=- L/2$  quantizes the single particle momenta $k_j$ which are expressed in terms of $M$ parameters $\lambda_\beta$, the Bethe rapidities or Bethe roots, which  satisfy a set of coupled nonlinear equations called the Bethe equations. In a state, $M$ denotes the number of down spins and $N-M$ is the number of up spins and vice-versa. We use the method of Boundary Algebraic Bethe Ansatz to obtain

\begin{eqnarray}\label{energy}
e^{-ik_jL}\!=\!\prod_{\alpha=1}^Mf(2b, 2\lambda_\alpha),~
f(x,z)=\!\prod_{\sigma=\pm}\frac{x+\sigma z+i}{x+\sigma z-i}
\end{eqnarray}
where $\lambda_\alpha$, $\alpha=1,\dots,M$ satisfy the Bethe Ansatz equations
\begin{eqnarray}\label{BAE}
\left[f(2\lambda_\alpha,2b)\right]^{N}f(2\lambda_{\alpha},2d)=\prod_{\alpha\neq \beta }^Mf(\lambda_\alpha,\lambda_\beta),
\end{eqnarray}

with $d=\sqrt{b^2-2b/c-1}$.
These Bethe equations correspond to Bethe reference state with all up and all down spins. The Bethe roots govern the spin degrees of freedom of the system and $M\leq N/2$ gives the total $z$-component of spin, $S^z=N/2-M$. The solutions to equations of type \eqref{BAE} are well studied in the literature \cite{Takahashi},\cite{ODBA}. The solutions $\lambda_\alpha$ can be real or take complex values in the form of strings. In order to have a non vanishing wavefunction they must all be distinct, $\lambda_\alpha \neq \lambda_\beta$. In addition, the value $\lambda_\alpha=0$ should also be discarded as it results in a vanishing wavefunction \cite{ODBA}. Bethe equations of the type \eqref{BAE} are reflective symmetric, that is they are invariant under $\lambda_\alpha\rightarrow -\lambda_\alpha$ transformation. Due to this symmetry, solutions to the Bethe equations occur in pairs $\{-\lambda_\alpha,\lambda_\alpha\}$. 

\smallskip

 Applying logarithm to \eqref{BAE} we obtain 

\begin{eqnarray}\nonumber
-\pi I_j+\sum_{\sigma=\pm} N\Theta(\lambda_\alpha+\sigma b,1/2)+\Theta(\lambda_\alpha+\sigma d ,1/2)\\+\Theta(\lambda_\alpha,1/2)+ \Theta(\lambda_\alpha,1)\label{Logbae}
=\sum_{\beta=-M}^{M} \Theta\left(\lambda_\alpha- \lambda_\beta,1\right)\end{eqnarray}

Where $\Theta(x,n)=\text{arctan}[x/n]$. And likewise taking the logarithm of \eqref{energy} we get
\begin{eqnarray}\label{logEnergy}
k_j=\frac{2\pi n_j}{L}+\frac{2}{L}\sum_{\beta=-M}^M\Theta( b -\lambda_\beta,1/2)
\end{eqnarray}

The integers $n_j$ and $I_\alpha$ arise from the logarithmic branch and serve as the quantum numbers of the states. The quantum numbers $I_\alpha$ correspond to the spin degrees of freedom while the quantum numbers $n_j$ are associated with the charge degrees of freedom and they must all be different. $I_\alpha$ and $n_j$ can be chosen independently implying the charge spin decoupling. Minimizing the ground state energy results in a cutoff such that $\pi|n_j|/L < \pi D$ where $D=N/L$ is the density \cite{AndreiLowenstein79}.  
We consider here $b>0$. The model exhibits several regimes depending on the values of $b$ and $c$ or equivalently $d$, where $d$ can be real or purely imaginary. Below we will solve the Bethe equations and construct the ground state separately in each regime.

\subsection{Ground state properties}
\label{BethesolutionAp}

\subsubsection{Kondo regime} 
The Kondo regime corresponds to all real values of $d$ and imaginary values $a<\frac{1}{2}$ where $d=ia$. Consider first the case where $d$ is real.  The ground state is given by the particular choice of charge and spin quantum numbers $n^0_j$ , $I^0_\alpha$, where $n^0_j$ are consecutively filled from the lower cutoff $-LD$ upwards, and the integers $I^0_\alpha$  take consecutive values which corresponds to real valued  $\lambda_\alpha$ roots. In the limit $N\rightarrow \infty$ the Bethe roots fill the real line and the ground state can be described by $\rho(\lambda)$ the density of solutions $\lambda$, from which the properties of the ground state can be obtained. Reflection symmetry of the Bethe equations \eqref{BAE} allows us to define $\lambda_{-\alpha}=-\lambda_\alpha, \; \lambda_0=0$ \cite{XXXkondo} and introduce the counting function $\nu(\lambda)$ such that $\nu(\lambda_\alpha)=I_\alpha$.

Differentiating \eqref{Logbae}, and noticing that $\rho(\lambda)=\frac{d}{d\lambda}\nu(\lambda)$ \cite{trieste}, we obtain the following integral equation, 

\begin{eqnarray}\label{gsrdensity}
g_{sr}(\lambda)&=&\rho_\text{sr}(\lambda)+\int_{-\infty}^{\infty}\mathrm{d}\mu\,\varphi(\lambda-\mu,1)\rho_\text{s}(\mu)
\end{eqnarray}

where $g_{sr}(\lambda)=\sum_{\sigma=\pm}N\varphi(\lambda+\sigma b ,1/2) +\varphi(\lambda+\sigma d ,1/2)+\varphi(\lambda,1/2)+\varphi(\lambda,1)$ and  $\varphi(x,n)= (n/\pi)(n^2+x^2)^{-1}.$

Solving \eqref{gsrdensity} by Fourier transform we obtain the following Fourier transformed density distribution of Bethe roots 

\bea \label{denscreened1}\tilde\rho_\text{sr}(\omega)= \frac{N_e \cos[ b\,\omega] + \cos[d \,\omega]+ \frac{1}{2}e^{-\frac{|\omega|}{2}}+\frac{1}{2}}{\sqrt{2\pi}\cosh[\frac{\omega}{2}]},\eea

 Each of the terms here may be identified with a certain component of the system. The term which is proportional to $N$ is the contribution of the left and right moving electrons, the next term which depends upon $d$ is the contribution due to the impurity while the remaining terms can be associated with the boundaries at $x=0,-L/2$.  The number of Bethe roots $M_{sr}$ in the ground state of Kondo regime  for $d$ real is given by
\bea
2M_{sr}+1=\int_{-\infty}^{+\infty}d\lambda \; \rho_{sr}(\lambda),
\label{nrootssr}
\eea
 from which the $z$-component of spin $(S^z)_{sr}$ of the ground state in this region is obtained using the relation $S^z_{sr}=N/2-M_{sr}$.  Taking into account that  $\tilde\rho(0)=\int\mathrm{d}\lambda\, \rho(\lambda)$ along with \eqref{denscreened1} we find that  \bea 
 (S^z)_{sr}=0.
 \eea 
 
Defining the fermionic parity as $\mathcal{P}=(-1)^{N_e}$ we find $\mathcal{P}=-1$. For the case of imaginary $d$ with $a<1/2$, $d=ia$, we have the following logarithmic form of the Bethe equations from \eqref{BAE}
 \begin{flalign}\nonumber
\sum_{\sigma=\pm} N\Theta(\lambda_\alpha+\sigma b,1/2)+\Theta(\lambda_\alpha,a+1/2)+\Theta(\lambda_\alpha,1/2-a)\\+\Theta(\lambda_\alpha,1) \label{Logbae2}
=\sum_{\beta=-M}^{M} \Theta\left(\lambda_\alpha- \lambda_\beta,1\right) +\pi I_j \end{flalign}

Differentiating \eqref{Logbae2} and following the same procedure as before we obtain

\begin{eqnarray}\label{gsidensity}
g_{si}(\lambda)&=&\rho_\text{si}(\lambda)+\int_{-\infty}^{\infty}\mathrm{d}\mu\,\varphi(\lambda-\mu,1)\rho_\text{s}(\mu)
\end{eqnarray}

where $g_{si}(\lambda)=\sum_{\sigma=\pm}N\varphi(\lambda+\sigma b ,1/2) +\varphi(\lambda, a+1/2)+\varphi(\lambda, 1/2-a)+\varphi(\lambda,1/2)+\varphi(\lambda,1)$.

Solving \eqref{gsidensity} by Fourier transform we obtain the following Fourier transformed density of roots

\bea  \label{denscreened2}\tilde\rho_\text{si}(\omega)= \frac{N_e \cos[ b\,\omega] + \cosh[a\;\omega]+ \frac{1}{2}e^{-\frac{|\omega|}{2}}+\frac{1}{2}}{\sqrt{2\pi}\cosh[\frac{\omega}{2}]},\eea

 The number of roots is given by formula same as \eqref{nrootssr}, from which we obtain the $z$-component of the spin of the ground state of Kondo regime for imaginary $d$ as \bea(S^z)_{si}=0.\eea Fermionic parity of this state is again $\mathcal{P}=-1$.  The ground state in the Kondo regime $\ket{0}$ is unique and is described by the distribution $\tilde\rho_\text{sr}(\omega)$ for d real and by the distribution $\tilde\rho_\text{si}(\omega)$ for imaginary $d$ with $a<1/2$, $d=ia$. Hence the ground state $\ket{0}$ in the Kondo regime has total spin $\vec{S}_T=0$ with odd fermionic parity. Therefore the impurity spin has been completely screened by the electrons in the bulk, indicative of the Kondo effect.
 
 \smallskip

 \subsubsection{YSR regime}
 The YSR regime corresponds to the range $1/2<a<3/2$. Consider the state with all real $\lambda_\alpha$. Applying logarithm to Bethe equations \eqref{BAE} we obtain
\begin{flalign} \label{Logbae3}\sum_{\sigma=\pm} N\Theta(\lambda_\alpha+\sigma b,1/2)+\Theta(\lambda_\alpha,a+1/2)-\Theta(\lambda_\alpha,a-1/2)\\+\Theta(\lambda_\alpha,1) 
=\sum_{\beta=-M}^{M} \Theta\left(\lambda_\alpha- \lambda_\beta,1\right) +\pi I_j \end{flalign}
 
 Differentiating \eqref{Logbae3} and following the same procedure as before we obtain

\begin{eqnarray}\label{gmdensity}
g_{us}(\lambda)&=&\rho_\text{us}(\lambda)+\int_{-\infty}^{\infty}\mathrm{d}\mu\,\varphi(\lambda-\mu,1)\rho_\text{s}(\mu)
\end{eqnarray}

where $g_{us}(\lambda)=\sum_{\sigma=\pm}N\varphi(\lambda+\sigma b ,1/2) +\varphi(\lambda, a+1/2)-\varphi(\lambda, a-1/2)+\varphi(\lambda,1/2)+\varphi(\lambda,1)$.

 Solving \eqref{gmdensity} by Fourier transform we obtain the following Fourier transformed density of roots

\bea  \label{denscreened3}\tilde\rho_\text{us}(\omega)= \frac{N_e \cos[ b\,\omega] + e^{-a|\omega|}-e^{-(a-1)|\omega|} +\frac{1}{2}e^{-\frac{|\omega|}{2}}+\frac{1}{2}}{\sqrt{2\pi}\cosh[\frac{\omega}{2}]},\eea
 
 The number of roots is given by formula same as \eqref{nrootssr}, from which we obtain the $z$-component of the spin of this state as \bea(S^z)_{us}=\frac{1}{2}.\eea 
  
This state has fermionic parity $\mathcal{P}=-1$.  Due to the $SU(2)$ symmetry we immediately deduce that there is another ground state in the same fermion parity sector, degenerate with the above, which has the opposite spin \bea
(S^z)_{\widehat{us}}=-\frac{1}{2}.
\eea

Actually, this state can be obtained by choosing the Bethe reference state with all spins down instead of up \cite{korepin1993quantum}. Hence there exists two degenerate states $\ket{1/2}$, $\ket{-1/2}$ in the YSR regime with spins $1/2$ and $-1/2$ respectively which are both described by the distribution $\tilde\rho_\text{us}(\omega)$. 

\smallskip

We now show that there exists another state $\ket{0'}$ in the YSR regime. This state can be obtained by adding a boundary string to either of the states $\ket{1/2}$, $\ket{-1/2}$. The boundary strings arise as purely imaginary solutions of the Bethe equations. These purely imaginary Bethe roots, which correspond to the bound states, appear as poles in the dressed or physical boundary S-matrix  \cite{gosh,gosz,skorik}. We categorize the boundary strings as short boundary strings and wide boundary strings if the absolute value of the imaginary part is lesser or greater than one respectively. By observation we see that, in the limit $N\rightarrow\infty$, for $a>1/2$, the Bethe equations \eqref{BAE} have a {\it unique} solution 
\be
\lambda_{bs}=\pm i(a-1/2),
\label{boundstring}
\ee 
as the two $\pm$ strings  leads to the same state by reflection symmetry.  $|Im(\lambda_{bs})|<1$ for $1/2<a<3/2$, hence it is a short boundary string in the YSR regime.

\smallskip 

Adding the boundary string \eqref{boundstring} to the Bethe equations \eqref{BAE} for $1/2<a<3/2$ and taking the logarithm we obtain 

\begin{flalign} \nonumber\sum_{\sigma=\pm} N\Theta(\lambda_\alpha+\sigma b,1/2)-\Theta(\lambda_\alpha,a-1/2)+\Theta(\lambda_\alpha,1)\\ \label{Logbae4}=\Theta(\lambda_\alpha,3/2-a)+\sum_{\beta=-M}^{M} \Theta\left(\lambda_\alpha- \lambda_\beta,1\right) +\pi I_j \end{flalign}
 
The above equation can be solved by following the same procedure as above, we obtain
  \bea
 \tilde{\rho}^b_{bs}(\omega)=\tilde{\rho}_{us}(\omega)+ \Delta\tilde{\rho}^b_{bs}(\omega), 
 \eea
where the shift $\Delta\tilde{\rho}^b_{bs}(\omega)$ is due to the presence of the boundary string which is given by
\bea  \label{BetherootsBS} \Delta\tilde{\rho}^b_{bs}(\omega)= \frac{e^{-(1-a)|\omega|}+e^{-a|\omega|}}{2\sqrt{2\pi}\cosh[\omega/2]} \eea

In the presence of  the boundary string, the relation between the number of Bethe roots and the density distribution also  takes a different form as compared to \eqref{nrootssr}. Namely
\bea 
\label{nrootsbs}2M^b_{bs}-1=\int_{-\infty}^{+\infty} d\lambda\; \rho^b_{bs}(\lambda),
\eea
from which, using $(S^z)^b_{bs}=N/2-M^b_{bs}$, we find
\bea 
(S^z)^b_{bs}=0. 
\eea 
Thus the resulting state $\ket{0'}$ described by the Bethe root distribution $\tilde{\rho}^b_{bs}(\omega)$ which has a boundary bound state has total spin $\vec{S}_T=0$ and has odd fermionic parity $\mathcal{P}=-1$.

To get the energy of this state, or of the boundary string, we notice that it is given 
by the energy difference, up to chemical potential, between  the ground states with $S^z=0$ and $S^z=\pm 1/2$
\begin{eqnarray}
\label{ebk}E_B=E^0_{N}-\frac{1}{2}(E^0_{N-1}+E^0_{N+1}).
\end{eqnarray}
Here $E^0_{N}$ refers to the energy of the state with odd number of particles which, in our system, corresponds to the state $\ket{0'}$ which includes the boundary string and has spin $S^z=0$. Similarly $E^0_{N+1}$ and  $E^0_{N-1}$ refer to the energies of the states $\ket{1/2}$ or $\ket{-1/2}$ with an {\it even} number of particles and spin $S^z=\pm 1/2$. The expression \eqref{ebk} is defined in \cite{Keselman2015} as the binding energy, which precisely measures the energy cost of adding an electron to the system. To calculate $E_B$ we use \eqref{logEnergy}, from which we obtain the following expression for total energy of a state with $N$  fermions
 \bea
 E=\sum_{j=1}^{N}\frac{\pi}{L}n_j+\frac{iD}{2}
 \int_{-\infty}^{\infty}d\lambda\; \Theta\left(b-\lambda,1/2\right)\rho(\lambda). \nonumber \\
 \label{toten}
 \eea
  From  \eqref{ebk} we find that $E_B$ has two contributions, one from the charge degrees of freedom 
 and one from the spin degrees of freedom: 
 $E_B=E_{\text{charge}}+\epsilon$. The charge contribution is given by the charging energy
 \begin{eqnarray}
  E_{\text{charge}}=\sum_{j=1}^{N} \frac{\pi}{L}n_j-\frac{1}{2}\left(\sum_{j=1}^{N+1} \frac{\pi}{L}n_j+\sum_{j=1}^{N-1} \frac{\pi}{L}n_j\right). \nonumber \\
  \label{totenc}
 \end{eqnarray}
 Note that the the charge quantum numbers take all the values from the cutoff $-DL$ upwards. In the ground state they fill all the slots from $n_j=-N \; \text{to}\; n_j=-1$. In the state with one extra particle they fill all the slots from $n_j=-N \;\text{to}\; n_j=0$. In the state with one less particle there is an unfilled slot at $n_j=-1$ which corresponds to a holon excitation. We obtain
 \bea \label{bounden}E_{\text{charge}}=-\frac{\pi}{2L},\eea
 
 hence it vanishes in the thermodynamical limit. The spin contribution is given by the expression
 \bea
 \epsilon=E_{0}+\frac{iD}{2}\int_{-\infty}^{+\infty} d\lambda\; \Theta\left(b-\lambda,1/2\right)\Delta\rho^b_{bs}(\lambda), \nonumber \\
 \label{totens}
 \eea

where $E_{0}=D\Theta(b-i(a-1/2),1/2)+D\Theta(b+i(a-1/2),1/2)$ and $\Delta\rho^b_{bs}(\lambda)$ is the shift of the Bethe roots distribution due  to the boundary string which is given in (\ref{BetherootsBS}). Evaluating (\ref{totens})  we find that 
 the spin part of the energy of the boundary string is

\bea \label{midgapen} \epsilon=-\Delta\sin(a\pi).\eea

Hence the energy of the boundary string or equivalently the energy difference between the states $\ket{0'}$ and $\ket{\pm1/2}$ is always less than the bulk gap $\Delta$. For $1/2<a<1$, $\epsilon<0$, hence the ground state is $\ket{0'}$ which contains a boundary mode which is bound to the edge where the impurity lives and screens it. Removing this boundary mode would un-screen the impurity which would cost energy less than the bulk gap yielding the states $\ket{\pm1/2}$. For $1<a<3/2$, $\epsilon>0$, hence the ground state comprises of the states $\ket{\pm1/2}$ and is two-fold degenerate. At $a=1$, $\epsilon=0$ and hence the ground state is three-fold degenerate where a level crossing occurs between the states $\ket{0'}$ and $\ket{\pm1/2}$. We also find that the energy difference between the ground state $\ket{0}$ in the Kondo regime and the ground state $\ket{0'}$ in the YSR regime vanishes as one approaches the boundary between these phases at $a=1/2$. 

\subsubsection{Unscreened regime}
The unscreened regime corresponds to $a>3/2$. By considering the state with all real Bethe roots $\lambda_\alpha$, one again obtains the states $\ket{1/2}$ and $\ket{-1/2}$ which are given by the distribution $\tilde{\rho}_{us}(\omega)$. The boundary string solution $\lambda_{bs}=i(a-1/2)$ still exists but adding this solution to the states $\ket{1/2}$ or $\ket{-1/2}$ is not possible unless one adds bulk holes. This is due to the fact that $|Im(\lambda_{bs})|>1$ for $a>3/2$, making it a wide boundary string.  The energy of the state corresponding to the addition of the wide boundary string goes above the mass gap due to the presence of the bulk hole and hence it does not correspond to a mid-gap state. Hence in the unscreened regime the ground state comprises of two states  $\ket{1/2}$ and $\ket{-1/2}$ and is two-fold degenerate.

\end{widetext}

\end{document}